\documentclass[twocolumn]{aastex631}

\usepackage{graphicx}	% Including figure files
\usepackage{amsmath}	% Advanced maths commands
\usepackage{amssymb}	% Extra maths symbols
\usepackage{newtxtext,newtxmath}
\usepackage{hyperref}

\newcommand{\RCR}{\ensuremath{R_{\textrm{CR}}}}
\newcommand{\Rot}{\ensuremath{\mathcal{R}}}
\newcommand{\Vc}{\ensuremath{V_{\textrm{c}} }}
\newcommand{\PS}{\ensuremath{\Omega_{\textrm{p}}}}
\newcommand{\Rb}{\ensuremath{R_{\textrm{b}}}}
\newcommand{\kms}{\ensuremath{\textrm{km}/\textrm{s}}}

\newcommand{\Nbody}{$N$-body}
\newcommand{\AREPO}{\textsc{AREPO}}
\newcommand{\SMUGGLE}{SMUGGLE}
\newcommand{\Msun}{\ensuremath{M_{\odot}}}

\shorttitle{Stellar Bars and Gas}
\shortauthors{Beane et al.}

\graphicspath{{./}{figures/}}

\begin{document}
\title{Stellar Bars in Isolated Gas-Rich Spiral Galaxies Do Not Slow Down}

\author{Angus Beane}
\affiliation{Center for Astrophysics $|$ Harvard \& Smithsonian,  Cambridge, MA, USA}

\author{Lars Hernquist}
\affiliation{Center for Astrophysics $|$ Harvard \& Smithsonian,  Cambridge, MA, USA}

\author{Elena D'Onghia}
\affiliation{Department of Physics, University of Wisconsin-Madison, Madison, WI, USA}
\affiliation{Department of Astronomy, University of Wisconsin-Madison, Madison, WI, USA}

\author{Federico Marinacci}
\affiliation{Department of Physics \& Astronomy `Augusto Righi', 
University of Bologna, Bologna, Italy}

\author{Charlie Conroy}
\affiliation{Center for Astrophysics $|$ Harvard \& Smithsonian,  Cambridge, MA, USA}

\author{Jia Qi}
\affiliation{Department of Astronomy, University of Florida, Gainesville, FL, USA}

\author{Laura V. Sales}
\affiliation{Department of Physics \& Astronomy, University of California, Riverside, CA, USA}

\author{Paul Torrey}
\affiliation{Department of Astronomy, University of Florida, Gainesville, FL, USA}

\author{Mark Vogelsberger}
\affiliation{Department of Physics, Massachusetts Institute of Technology, Cambridge, MA, USA}

\begin{abstract}

  Elongated bar-like features are ubiquitous in galaxies, occurring at the
  centers of approximately two-thirds of spiral disks in the nearby Universe.  
  Due to gravitational interactions between the bar and the other components of
  galaxies, it is expected that angular momentum and matter will redistribute
  over long (Gyr) timescales in barred galaxies. Previous work ignoring the gas
  phase of galaxies has conclusively demonstrated that bars should slow their
  rotation over time due to their interaction with dark matter halos. We have
  performed a simulation of a Milky Way-like galactic disk hosting a strong bar
  which includes a state-of-the-art model of the interstellar medium and a live
  dark matter halo. In this simulation the bar pattern does not slow down over
  time, and instead remains at a stable, constant rate of rotation. This
  behavior has been observed in previous simulations using more simplified
  models for the interstellar gas, but the apparent lack of secular evolution
  has remained unexplained. We find that the presence of the gas phase
  arrests the process by which the dark matter halo slows down a bar, a
  phenomenon we term bar locking. This locking is responsible for stabilizing
  the bar pattern speed. We find that in a Milky Way-like disk, a gas fraction
  of only about 5\% is necessary for this mechanism to operate. Our result
  naturally explains why nearly all observed bars rotate rapidly and is
  especially relevant for our understanding of how the Milky Way arrived at its
  present state.

\end{abstract}

%% Keywords should appear after the \end{abstract} command. 
%% The AAS Journals now uses Unified Astronomy Thesaurus concepts:
%% https://astrothesaurus.org
%% You will be asked to selected these concepts during the submission process
%% but this old "keyword" functionality is maintained in case authors want
%% to include these concepts in their preprints.

%% From the front matter, we move on to the body of the paper.
%% Sections are demarcated by \section and \subsection, respectively.
%% Observe the use of the LaTeX \label
%% command after the \subsection to give a symbolic KEY to the
%% subsection for cross-referencing in a \ref command.
%% You can use LaTeX's \ref and \label commands to keep track of
%% cross-references to sections, equations, tables, and figures.
%% That way, if you change the order of any elements, LaTeX will
%% automatically renumber them.
%%
%% We recommend that authors also use the natbib \citep
%% and \citet commands to identify citations.  The citations are
%% tied to the reference list via symbolic KEYs. The KEY corresponds
%% to the KEY in the \bibitem in the reference list below. 

\section{Introduction}
\label{sec:intro}
\begin{figure*}
    \centering
    \includegraphics[width=\textwidth]{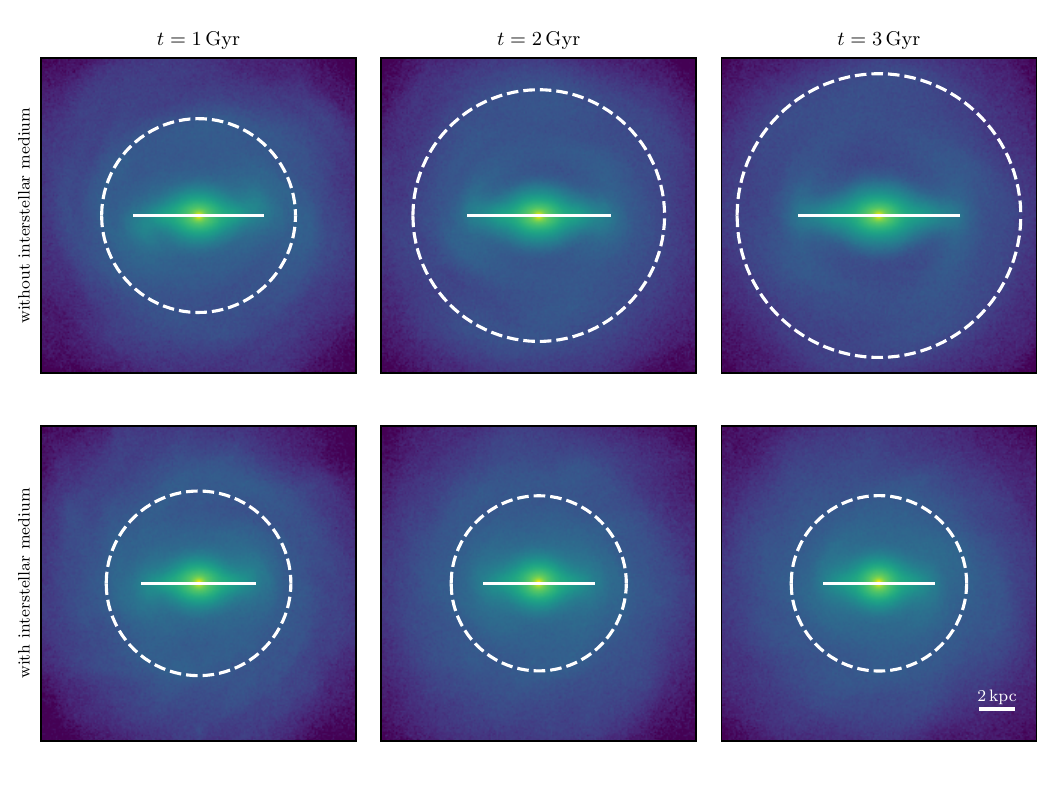}
    \caption{Stellar mass distribution of our simulations with and without
    the interstellar medium. The upper panels show an \Nbody{} only simulation
    while the lower panels show a simulation which includes the \SMUGGLE{} model
    for the interstellar medium. Each panel is $20\,\textrm{kpc}$ to a side.
    The bar length is shown as a solid white bar (details on its
    computation are given in the text). The corotation radius is shown as a
    dashed white circle. Columns show different points in time, separated by
    $1\,\textrm{Gyr}$. We can see that in the \Nbody{} run, the bar grows in
    length and strength. In the \SMUGGLE{} run, the bar remains at approximately
    the same length and strength over the course of the simulation. The \Nbody{}
    model is identical to the GALAKOS model (with different particle
    numbers and softening lengths), discussed in the text.}\label{fig:overview}
\end{figure*}

Approximately two-thirds of spiral disks host an elongated bar-like feature at
their centers \citep{2000AJ....119..536E, 2007ApJ...657..790M}, including our
own Milky Way \citep{1957AJ.....62...19J, 1991ApJ...379..631B}. The ubiquity of
bars is not difficult to explain, since stellar disks simulated in isolation
almost always form bar-like structures \citep{1971ApJ...168..343H}. Several
studies have shown that a hot, centrally concentrated mass distribution,
such as a stellar bulge or dark matter halo, acts to stabilize stellar disks
against bar formation \citep[e.g.,][]{1973ApJ...186..467O, 1976AJ.....81...30H}.

It is more difficult to reconcile numerical simulations with the observed
pattern speeds of extragalactic bars. Currently, the best technique for
measuring the pattern speeds of individual galaxies is the Tremaine-Weinberg
method \citep{1984ApJ...282L...5T, 2011MSAIS..18...23C}. This approach has
recently been applied to samples of galaxies from the MaNGA survey
\citep{2019MNRAS.482.1733G, 2020MNRAS.491.3655G} and CALIFA survey
\citep{2015AA...576A.102A}. These studies confirm what was found in earlier
works, that nearly all extragalactic bars are fast rotators (i.e., they rotate
close to their maximum rotation rate).

This is a problem for theoretical simulations, for which there is ample evidence
that galactic bars should resonantly interact with the dark matter halo, causing
the bar to slow down over time \citep{1981AA....96..164C, 1992ApJ...400...80H,
2000ApJ...543..704D, 2002MNRAS.330...35A, 2002ApJ...569L..83A,
2003MNRAS.341.1179A, 2003MNRAS.346..251O, 2005MNRAS.363..991H,
2006ApJ...637..214M, 2007MNRAS.375..460W, 2009ApJ...697..293D}. The physical
mechanism of this interaction can be understood as an angular form of dynamical
friction between the bar and the dark matter halo. While studied in detail for
the bar \citep{1984MNRAS.209..729T, 1985MNRAS.213..451W}, this process is
generic for any non-axisymmetric disturbance \citep{1972MNRAS.157....1L}. (For
an old but still useful review of bar dynamics, see
\citet{1993RPPh...56..173S}.)

Bar pattern speeds are usually measured using the parameter $\Rot=\RCR/\Rb$,
where \RCR{} is the corotation radius and \Rb{} is the bar length.\footnote{The
radius of corotation \RCR{} is defined for circular orbits as the radius at
which the orbital frequency is equal to the pattern speed, \PS{}, of a given
non-axisymmetric feature. In a galaxy with a constant circular velocity \Vc{},
it is given by $\RCR = \Vc / \PS$.} Bars with $\Rot < 1.4$ are considered
``fast rotators'' while bars with $\Rot > 1.4$ are considered ``slow
rotators'' \citep{2000ApJ...543..704D}. Bars with $\Rot < 1$ are not
thought to be stable \citep{1980AA....81..198C}. Observational estimates of the
pattern speeds of bars indicate that nearly all galaxies have $1 < \Rot < 1.4$
\citep{2011MSAIS..18...23C, 2015AA...576A.102A, 2019MNRAS.482.1733G,
2020MNRAS.491.3655G}. We note that \citet{2017ApJ...835..279F} argue that the
pattern speed should be measured relative to a characteristic angular velocity
of the outer disk.

While the fact that a bar is slowed down by a dark matter halo is
well-understood theoretically, this is not the case for the interaction between
a bar and a gaseous disk. Some argue that the gas disk should slow down the bar
more \citep{2003MNRAS.341.1179A}, while others argue that the tendency of the
bar to drive gas inwards means the bar should speed up due to the effect of the
gas disk \citep{2013MNRAS.429.1949A, 2014MNRAS.438L..81A}. Since the gas phase
typically contributes only about $10-20\%$ of the mass of a galaxy at the
present day, one might naively expect it to have a subdominant effect on the
bar. However, because gas is collisional, it can participate in non-resonant
angular momentum exchange with the bar \citep{2011MNRAS.415.1027H}. Thus,
numerical work has shown that the gas phase can have a stronger influence on a
bar than its contribution to the mass of a galaxy would suggest
\citep[e.g.,][]{2010ApJ...719.1470V, 2013MNRAS.429.1949A}.

In the last decade, stellar bars have been studied mainly in the context of the
instability processes that lead to their formation and their ability to drive
gas toward the galaxy center and contribute to the formation of supermassive
black holes (SMBH). The primary interest of these studies mainly was in the loss
of angular momentum of the gas and the associated  gaseous flow down to the
inner disk, possibly forming a central mass concentration
\citep{2010ApJ...719.1470V}, and fueling the central SMBH
\citep[e.g.][]{1989Natur.338...45S, 1990Natur.345..679S}. A revisit of the more
general problem of galactic bar properties and formation, including the case of
disk galaxies with very large gas fraction is timely since it is clear now that
galactic disks show massive bars already at redshift $z>2$
\citep{2022arXiv221008658G}. At that time the universe was 2.5 billion year old
and galaxies might have as high gas fraction as 80\% \citep{2020ARAA..58..157T}.
Furthermore, unlike nearby disk galaxies, the high-redshift disks also
continuously accrete cold gas from the cosmic web, making the formation,
stability and evolution of non-axisymmetric features a key question to address
since they can play a fundamental role in the more general problem of disk
galaxy evolution.

We have performed a simulation of a disk galaxy using the finite-volume,
gravito-hydrodynamics code \AREPO{} \citep{2010MNRAS.401..791S}. We use the
galaxy formation model Stars and MUltiphase Gas in GaLaxiEs
\citep[\SMUGGLE{};][]{2019MNRAS.489.4233M}. This disk galaxy exhibits almost no
evolution in the bar pattern speed over several Gyr when the gas phase is
accounted for and robustly modeled. This behavior has been observed in
previous works \citep{1993AA...268...65F, 2007ApJ...666..189B,
2009ApJ...707..218V, 2010ApJ...719.1470V, 2014MNRAS.438L..81A}. We propose a new
physical explanation for this stable behavior.

In particular, we find that the presence of a gas phase in a barred
galaxy can arrest the process by which the dark matter halo brakes the bar. The
constant positive torque on the bar by the gas causes the bar's resonance to
reside in regions which have become depopulated. In the halo wake picture, this
is equivalent to no new material being available to reinforce the wake. We show
that this occurs in a Milky Way-like disk with gas fractions as low as
about $5\%$.

We show stellar mass distributions of our barred galaxy in a case with and
without gas in Fig.~\ref{fig:overview}. We see that the bar grows longer and
stronger without gas (bar length shown as a white bar), while it remains
at approximately the same length and strength when gas is included. As the
bar without gas slows down, the corotation radius (white dashed circle) grows
larger with time.

% In this work, we present a simulation of an isolated, gas-rich disk galaxy. Our
% galaxy exhibits some similarities to the Milky Way, described in more detail in
% Section XX. Surprisingly, we find that the pattern speed of the bar in this
% galaxy is constant with time. We find that with even modest gas fractions of
% about 5\%, our isolated galaxy displays no evolution in pattern speed over many
% $\textrm{Gyr}$. 

In Section~\ref{sec:methods}, we describe our initial setup, numerical model,
and details on our bar analysis procedures. In Section~\ref{sec:results}, we
summarize the main results from our findings. We discuss these findings at more
length and in the context of previous research in Section~\ref{sec:discussion}
before concluding in Section~\ref{sec:conclusions}.

\section{Methods}
\label{sec:methods}
\begin{figure}
    \centering
    \includegraphics[width=\columnwidth]{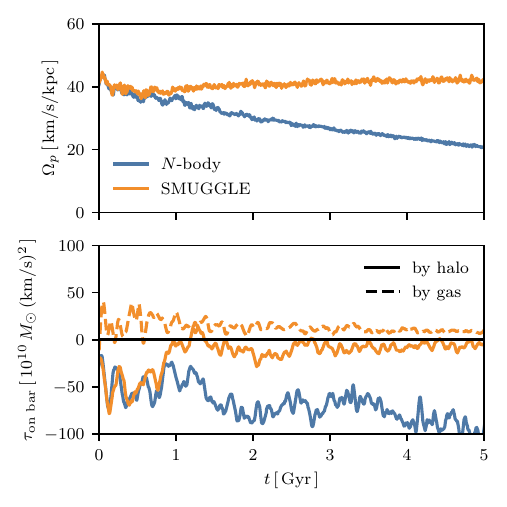}
    \caption{The \textit{upper panel} shows the evolution of the pattern speed.
    As expected, the bar in the \Nbody{} run slows down due to interactions
    between the bar and the dark matter halo. However, the bar in the \SMUGGLE{}
    run does not slow down and instead remains at a constant pattern speed. The
    \textit{lower panel} shows the torque on the bar by different components.
    The solid lines indicate the torque exerted by the halo in both the \Nbody{}
    and \SMUGGLE{} cases. The dashed line is the torque exerted by the gas phase
    in the \SMUGGLE{} run (there is no gas in the \Nbody{} run). After
    $\sim1\,\textrm{Gyr}$ of evolution, the torque by the halo in the \SMUGGLE{}
    case is severely reduced. We call this bar locking, and discuss its proposed
    origin in Section~\ref{sec:discussion}. Details on the calculation of the
    torque and pattern speed is given in
    Section~\ref{ssec:bar_analysis}.}\label{fig:prop}
\end{figure}

\begin{figure}
    \centering
    \includegraphics[width=9cm]{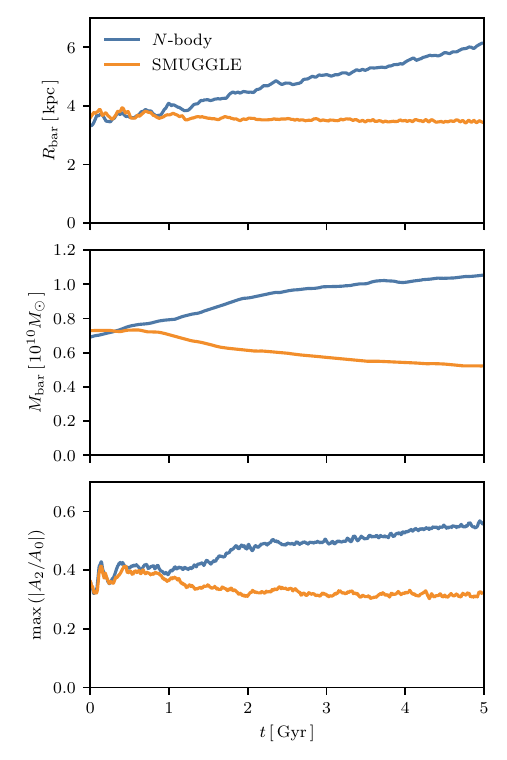}
    \caption{The evolution in bar length, mass and strength. The \textit{upper panel}
    shows the evolution of the bar length. In the \Nbody{} case, the bar
    lengthens. This occurs because as the pattern speed drops, bar-like orbits
    at larger radii are possible. Stars are captured on these orbits,
    lengthening the bar. This process does not occur in the \SMUGGLE{} cases
    since the bar pattern speed is not decreasing, and therefore the bar length
    remains constant. The bar mass, shown in the \textit{middle panel}, is
    increasing in the \Nbody{} case as the bar grows. It is decreasing in the
    \SMUGGLE{} case, indicating mass redistribution without a change in
    pattern speed. The bar strength, shown in the \textit{lower panel} is
    measured as the maximum of the second Fourier component divided by the
    zeroth Fourier component. We see that in the \Nbody{} case (blue) the bar
    strength increases with time, consistent with previous results showing that
    the bar strength increases as bars slow down. In the \SMUGGLE{} case
    (orange), we see that the bar strength slightly decreasing with time. This
    is also consistent with the expected relation between pattern speed and
    strength since the bar in this case is not slowing down.}
    \label{fig:strength}
\end{figure}
\subsection{Initial Conditions}
The initial setup of the galactic disk used in this work follows closely the
GALAKOS model \citep{2020ApJ...890..117D}, which uses a modified version of the
\texttt{MakeNewDisk} code \citep{2005MNRAS.361..776S}. The GALAKOS model has
three components - a radially exponential and vertically isothermal stellar
disk, and a stellar bulge and dark matter halo following a Hernquist profile
\citep{1990ApJ...356..359H}. All \Nbody{} runs in this work used the same setup
parameters as the GALAKOS disk, more details of which can be found in the
original paper.

The addition of the gas phase was done as follows. The version of
\texttt{MakeNewDisk} used for the original GALAKOS model can generate a gas disk
which is radially exponential and in vertical gravito-hydrodynamic balance. We
modified the radial profile of this code in order to allow us to generate a disk
with a constant surface density within some cut-off radius, and then
exponentially declining beyond that radius with the scale-length of the stellar
disk. Our fiducial model used an initial surface density of
$20\,M_{\odot}/\textrm{pc}^2$ and a cut-off radius of $9.3\,\textrm{kpc}$.
This corresponds to an initial gas fraction of $\sim16\%$. The initial gas disk
is generated with a temperature of $10^4\,\textrm{K}$ and solar metallicity.

After generating the gaseous disk in this way, we stitched the gas disk together
with the GALAKOS \Nbody{} disk (and bulge and dark matter halo) after the
GALAKOS disk has been allowed to evolve for $1.5\,\textrm{Gyr}$. The purpose of
allowing the GALAKOS disk to evolve first for a short period of time is to allow
for the bar to form unimpacted by the presence of the gas. We found that
including the gas before the bar has formed disrupts the formation of the bar,
as has been seen previously \citep[e.g.,][]{2013MNRAS.429.1949A}. Throughout
this work, we consider $t=0$ for the \Nbody{} run to be the time at which we
added the gas phase for the \SMUGGLE{} run (i.e., we ignore the first
$1.5\,\textrm{Gyr}$ of evolution of the \Nbody{} disk when the bar is forming).

We made one additional modification when stitching the gas disk together with
the \Nbody{} disk - we created a hole within the central $4\,\textrm{kpc}$ of
the gas disks. This hole guards against an initial dramatic infall of gas within
the bar region, which we found to destroy the bar. It is not uncommon for
observed barred galaxies to have gas deficits in the bar region \citep[though
not in the very center;][]{1993RPPh...56..173S}. Therefore, our practice of
allowing the gas distribution to have a hole in the central region is consistent
with our choice to begin the simulations with a bar already formed. In this
manner, we are able to study the ensuing self-consistent interaction between the
bar and the gas, but of course we are unable to explore the origin of bars in
the presence of the gas.

Our method for initializing gas was arrived at after numerous attempts to
include enough gas in the simulation to be compatible with the Milky Way while
also not destroying the bar. For example, we tried evolving an exponential gas
disk adiabatically with the barred disk (i.e., no cooling, star formation, or
feedback). Because there is no mechanism to remove gas from the central region,
a large, highly pressurized pileup of gas forms in the center.\footnote{ We
noticed that the bar slows down in this adiabatic model which lacks a mechanism
to remove gas from the central region. \citet{2007ApJ...666..189B} also noted
slowdown behavior in models which lacked central gas removal. It appears that
gas removal from the center (in our model due to star formation) is necessary to
stabilize a bar's pattern speed.} We then turned on the full SMUGGLE model. As
a result, there is a sudden collapse of gas to the center as pressure support is
lost due to cooling and star formation. This abrupt change in the potential
destroys the bar. 

We don't believe our method is the only nor even the best way to include gas.
One advantage of our method, though, is that it approximates the expectation
that in the bar region the gas surface density should be significantly reduced
\citep[e.g.][]{1993RPPh...56..173S}

We used a mass resolution of $7.5\times10^3\,M_{\odot}$ for the baryonic
components (initial stellar disk, stellar bulge, and gas) and a mass resolution
of $3.75\times10^4\,M_{\odot}$ for the dark matter halo. This mass resolution is
closest to ``level 3'' in the AURIGA simulations \citep{2017MNRAS.467..179G}.
This corresponds to approximately $6.4\times10^6$ particles in the stellar disk,
$1.1\times10^6$ in the bulge, $1.2\times10^6$ in the gas disk, and
$25.3\times10^6$ in the dark matter halo. We used a softening length of
$20\,\textrm{pc}$ for all collisionless components. This softening length is
smaller than used in the original GALAKOS model ($28\,\textrm{pc}$ in their
model, but $\sim43.5\,\textrm{pc}$ when scaled to our mass resolution). Our
smaller softening length is consistent with other resolved ISM models
\citep{2018MNRAS.480..800H, 2019MNRAS.489.4233M}. However, as a consistency
check, we reran our fiducial SMUGGLE run with $40\,\textrm{pc}$ softening and
found no difference in the pattern speed evolution of the bar. For the gas
component, the softening length is fully adaptive with a softening factor of
$2.5$ \citep[e.g.,][]{2020ApJS..248...32W}. Snapshots were saved at equal
intervals of $0.005$ in the time units of the simulation,
$\textrm{kpc}/(\textrm{km}/\textrm{s})$.

Our setup is initially out of equilibrium, but we found that after about
$500\,\textrm{Myr}$, the system has settled into a roughly steady-state
configuration and initial transients appear not to affect the results after this
point. The constant surface density of the initial gas disk is important for
ensuring that the gas disk is dense enough in order for comparisons to real
galaxies to be appropriate.

\subsection{Numerical Model}
We use the Stars and MUltiphase Gas in GaLaxiEs (\SMUGGLE{}) model
\citep{2019MNRAS.489.4233M} implemented within the moving-mesh, finite-volume
hydrodynamics and gravity code \AREPO{} \citep{2010MNRAS.401..791S}. The \SMUGGLE{}
model additionally includes radiative heating and cooling, star formation, and
stellar feedback. Explicit gas cooling and heating of the multi-phase
interstellar medium is implemented, covering temperature ranges between $10$ and
$10^8\,\textrm{K}$.

Star formation occurs in cells above a density threshold
($n_{\textrm{th}}=100\,\textrm{cm}^{-3}$) with a star-formation efficiency of
$\epsilon = 0.01$. Star formation converts gas cells into star particles which
represent single stellar populations with a Chabrier initial mass function
\citep{2003PASP..115..763C}. For each star particle, the deposition of energy,
momentum, mass, and metals from stellar winds and supernovae is modeled.
Photo-ionization and radiation pressure are handled using an approximate
treatment. A more detailed description of this model can be found in the
flagship \SMUGGLE{} paper \citep{2019MNRAS.489.4233M}. A pedagological
review of cosmological simulations of galaxy formation can be found in
\citet{2020NatRP...2...42V}.

We used the fiducial model parameters, except that we increased the number of
effective neighbors $N_{\textrm{ngb}}$ for the deposition of feedback from $64$
to $512$. We found that a lower value of $N_{\textrm{ngb}}$ resulted in
inefficient photo-ionization feedback since the photo-ionizing budget had not
been exhausted after deposition into $64$ neighboring cells. We also employed an
updated version of \SMUGGLE{} using a new mechanical feedback routine similar to
the one described in \citet{2018MNRAS.480..800H}. This updated routine is a
tensor renormalization which ensures linear and angular momentum conservation to
machine precision.

In addition to the \SMUGGLE{} model, we considered a simpler model of the
interstellar medium based upon \citet{2003MNRAS.339..289S}. In this approach,
the multiphase nature of the interstellar medium is described in a subgrid
manner by allowing each resolution element to have a ``cold'' and ``hot''
component, with the equation of state of the gas suitably modified. Gas is
allowed to interchange between the cold and hot components through processes
such as cooling and stellar feedback. Cold gas is allowed to undergo star
formation. We refer to this model as the smooth interstellar medium model, and
it is described in more detail in \citet{2019MNRAS.489.4233M}.

\subsection{Bar Analysis}
\label{ssec:bar_analysis}
The analysis of various bar properties is performed as follows. First, the
pattern speed is measured from the angle of the second Fourier component. We
measured the second Fourier component by computing,
\begin{equation}
\begin{split}
A_2 &= \sum_i m_i e^{i 2 \phi_i} \\
A_0 &= \sum_i m_i \textrm{,}
\end{split}
\end{equation}
where $m_i$ and $\phi_i$ are the mass and azimuthal angle of each particle,
respectively. We computed $A_2$ and $A_0$ in cylindrical bins of width
$0.5\,\textrm{kpc}$ from radii of $0$ to $30\,\textrm{kpc}$. We defined the
angle of the bar $\phi_b$ to be half the angle of the complex number $A_2$
as measured in the bin extending from a radius of $2.5$ to $3\,\textrm{kpc}$.
After correcting for the periodicity of $\phi_b$, we measured the pattern speed
as one-half the two-sided finite gradient of $\phi_b$ as a function of time.
We note that using the second Fourier mode is a blunt tool, discussed in
\citet{2019arXiv190308203P}.

In order to compute other properties of the bar, it is necessary to decompose
the disk into a barred and unbarred component. We achieved this by following
closely the methods described in \citet{2016MNRAS.463.1952P,
2021MNRAS.500..838P}. Our implementation is described in more detail in
Appendix~\ref{app:bardecomp}. After the disk has been decomposed into a trapped
and untrapped component, we measured the bar length as being the radius $R_b$
which encapsulates $99\%$ of the stars identified as being trapped in the bar.

To compute torques we used the tree algorithm in \texttt{MakeNewDisk}
\citep{2005MNRAS.361..776S} customized to be accessible from \texttt{Python}
using \texttt{Cython}. This algorithm is based on the \texttt{TREESPH} code
\citep{1989ApJS...70..419H}. We constructed a tree with an opening angle of
$0.35$ using only the star particles identified as being trapped in the bar. We
then queried the tree at the locations of all resolution elements in the other
components and computed the torque of the bar on such components. The torque on
the bar by the other components is simply the negative of the torque on the
other components by the bar. A similar analysis using basis function expansions
was performed in \citet{2019MNRAS.490.3616P}.

\subsection{Plotting Details}
We saved snapshots in intervals of $0.005$ in the time units of the simulation,
$\textrm{kpc}/(\textrm{km}/\textrm{s})$, which is very nearly equal to
$1\,\textrm{Gyr}$ (it is $\sim0.977\,\textrm{Gyr}$). Therefore, throughout this
work we referred to the native code time unit as $\textrm{Gyr}$. None of our
results are sensitive to this choice. We applied a Savitzky-Golay filter
\citep{1964AnaCh..36.1627S} as implemented in \texttt{scipy} using a window
length of $21$ and polynomial order of $3$ to the plot of torques
(Fig.~\ref{fig:prop} and Fig.~\ref{fig:sam-torque}) and angle differences
(Fig.~\ref{fig:wake}) in order to remove some numerical noise.

\begin{figure}
    \centering
    \includegraphics[width=\columnwidth]{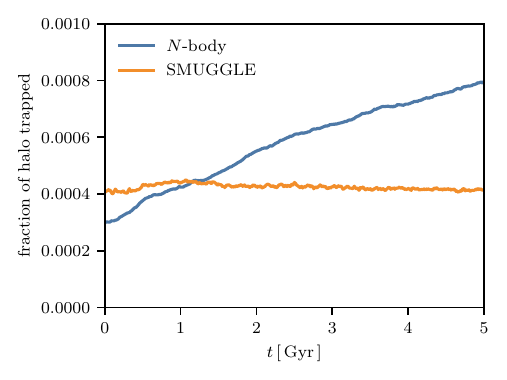}
    \caption{The halo mass fraction within two disk scale lengths
    ($\sim5.3\,\textrm{kpc}$) trapped in the \Nbody{} and \SMUGGLE{} runs. As
    the dark matter halo torques the bar, material from the halo is trapped on
    bar-like orbits. In the \Nbody{} case, the trapped fraction increases with
    time, indicating the torquing process is active and the bar is unlocked. In
    the \SMUGGLE{} case, the trapped fraction is nearly constant with time,
    indicating the torquing process is inactive and the bar is locked.}
    \label{fig:htrap}
\end{figure}

\begin{figure*}
    \centering
    \includegraphics[width=\textwidth]{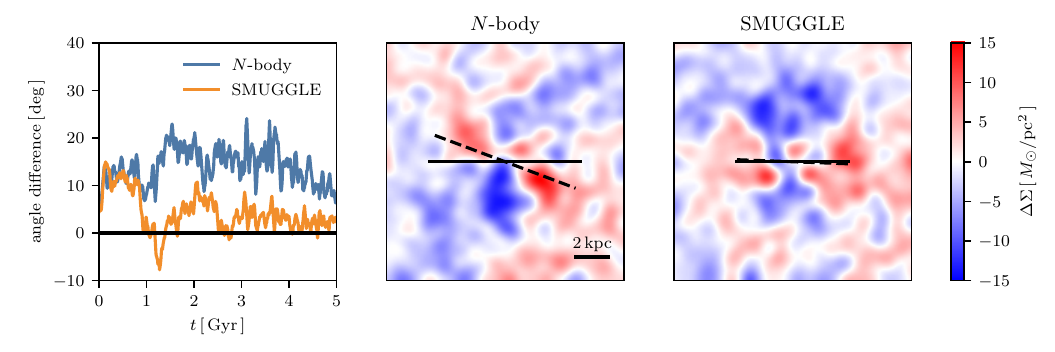}
    \caption{The wake excited in the dark matter halo. The dark matter halo wake
    is shown in the \Nbody{} case (\textit{middle panel}) and \SMUGGLE{} case
    (\textit{right panel}) after $2.6\,\textrm{Gyr}$ of evolution. The
    \textit{middle} and \textit{right panels} show a surface density projection
    in the $x$-$y$ plane of the dark matter halo after an axisymmetric average
    has been subtracted. The solid line indicates the direction of the bar while
    the dashed line indicates the direction of the halo wake (both measured by
    taking the second Fourier component within a sphere of all material within a
    radius of $4\,\textrm{kpc}$). The \textit{left panel} shows the time
    evolution of the angle difference between the bar and the halo wake, as
    measured from the second Fourier component. After the first Gyr, the angle
    difference in the \SMUGGLE{} case is smaller than in the \Nbody{} case by about
    a factor of two, reflecting how the dark matter halo in the \SMUGGLE{} case is
    unable to exert as negative a torque on the bar as in the \Nbody{} case.}
    \label{fig:wake}
\end{figure*}

\section{Results}
\label{sec:results}

We present the time evolution of different bar properties in Fig.~\ref{fig:prop}.
In the upper panel, we show the pattern speed over time in the \Nbody{} (blue)
and \SMUGGLE{} (orange) runs. The pattern speed in the \Nbody{} case slows down
while the pattern speed in the \SMUGGLE{} case remains roughly constant. The
slowing down of the pattern speed in the \Nbody{} case is consistent with a long
line of numerical research on bars in \Nbody{} simulations
\citep{1992ApJ...400...80H, 2000ApJ...543..704D, 2002MNRAS.330...35A,
2002ApJ...569L..83A, 2003MNRAS.341.1179A, 2003MNRAS.346..251O,
2005MNRAS.363..991H, 2006ApJ...637..214M, 2007MNRAS.375..460W,
2009ApJ...697..293D}.

However, in the \SMUGGLE{} case the pattern speed remains constant. After the first
Gyr of evolution, we find that the pattern speed increases by only $\sim10\%$
over the next $4\,\textrm{Gyr}$, compared to a $\sim43\%$ decrease in the
pattern speed for the \Nbody{} run over the same interval.

The bottom panel of Fig.~\ref{fig:prop} shows the torque exerted on the bar by
different components. The solid lines indicate the torque on the bar by the dark
matter halo, whereas the dashed line is the torque on the bar by the gas phase.
In the \Nbody{} case, the halo exerts a steady negative torque on the bar, with
an average torque from $1$ to $4\,\textrm{Gyr}$ of $-58.0$ in units of
$10^{10}M_{\odot}\,(\textrm{km}/\textrm{s})^2$. The halo in the \SMUGGLE{} case
exerts a similar negative torque on the bar in the first Gyr of evolution, but
after that the halo exerts a much weaker torque on the bar, averaging only
$-7.8$ in the same units and over the same time interval. The gas in the \SMUGGLE{}
case exerts a steady positive torque averaging $11.7$ over $1\,\textrm{Gyr}$ in
the same units.

As we saw qualitatively in Fig.~\ref{fig:overview}, the upper panel of
Fig.~\ref{fig:strength} shows that the length of the bar in the \Nbody{} case
grows over time while it remains roughly constant in the \SMUGGLE{} case. This
is also consistent with previous numerical work, which found that bars tend to
grow as they slow down and the radius of corotation increases
\citep{2000ApJ...543..704D, 2003MNRAS.341.1179A}. The middle panel of
Fig.~\ref{fig:strength} shows the mass of the bar. As the \Nbody{} bar slows
down and lengthens, it also grows in mass. The \SMUGGLE{} bar, however, loses
mass over time. This indicates a change in the bar's angular momentum without 
a change in pattern speed, and highlights the
fact that bars are not solid bodies and can respond to external torques through
mass redistribution.

The time evolution of the bar strength, defined as the maximum of
$\left|A_2/A_0\right|$ as a function of radius, is shown in the lower panel of
Fig.~\ref{fig:strength}. The quantity $\left|A_2/A_0\right|$ varies from $0$ to
$1$, with larger values indicating a stronger bar pattern. We see that in the
\Nbody{} case, $\left|A_2/A_0\right|$ increases over time as the bar pattern
slows. This is consistent with previous \Nbody{} simulations which showed a
clear correlation between the bar pattern speed and the bar strength
\citep[e.g.,][]{2003MNRAS.341.1179A}. In the \SMUGGLE{} case, we see that the
bar strength has an initial drop but then remains at a roughly constant, but
slightly decreasing, strength. This is consistent with the pattern speed in the
\SMUGGLE{} case being roughly constant or slightly increasing.

\section{Discussion}
\label{sec:discussion}
\subsection{Pattern Speed Evolution}
The lack of evolution in the pattern speed of the \SMUGGLE{} case (seen in
Fig.~\ref{fig:prop}) is intimately tied to the sudden decrease in torque exerted
on the bar by the dark matter halo. 

%We argue that this can be understood in terms of the halo wake mechanism. 
We interpret this behavior in terms of the halo wake mechanism. In the
\Nbody{} case, a well known phenomenon is that the halo material resonant with
the bar forms a wake, and this wake lags behind \citep{1984MNRAS.209..729T,
1985MNRAS.213..451W, 1992ApJ...400...80H} and exerts a negative torque on the
bar, slowing it down (see Fig.~\ref{fig:wake} below).\footnote{Since the bar is not a solid
body, it is not guaranteed that a negative torque will slow it down - e.g. a
negative torque could reduce the mass of the bar, reducing its moment of inertia without
changing its pattern speed. However, the bar seems to empirically respond to a
negative torque induced by a halo wake by slowing down.} As the bar slows down,
the location of the resonances in the phase space changes (see Fig.~12 and Table
1 in \citet{2020ApJ...890..117D}) allowing halo material newly resonant with the
bar to participate in the formation of the wake. However, the gas is also a
reliable source of positive torque on the bar, speeding the bar up. In turn,
this stops the location of the resonance from changing such that the halo cannot
reinforce the wake, therefore arresting the process by which the halo can slow
the bar down. We term this process ``bar pattern speed locking,'' or simply
bar locking for short.\footnote{ To be more explicit, it is the gas which locks 
the pattern speed of the bar. The gas does this by forcing the resonant locations 
of the bar into regions of the halo phase space that can no longer support 
significant negative torque.} 

This bar locking process is similar to the ``metastability'' effect, which has
been previously discussed in the literature \citep{2003MNRAS.345..406V,
2006ApJ...639..868S}. Finally, we also note that these authors, in particular,
observe that the effects of numerical resolution in the simulations adopted to
explore these mechanisms have yet to be fully explored and could play a role in
the observed phenomenology of the simulations. We plan to address these issues
in future dedicated work.

We test this interpretation in two ways. First, we measure the fraction of
mass trapped in the halo. As material in the halo wake torques the bar, that
material becomes trapped on bar-like orbits \citep[the ``shadow
bar'';][]{2016MNRAS.463.1952P}. In Fig.~\ref{fig:htrap} we show the halo trapped
fraction for particles with radii less than two disk scale lengths
($\sim5.3\,\textrm{kpc}$). In the \Nbody{} case, the trapped fraction increases
with time, as expected since the halo is actively torquing the bar (which is,
therefore, unlocked). In the \SMUGGLE{} case, the trapped fraction is constant
(or perhaps slightly decreasing) with time (indicating the bar is locked). This
supports our interpertation that the halo wake process has shut down in the
presence of the gas phase.

Second, we measure the angle offset between the halo wake and the bar. If the
wake and the bar are aligned (i.e., there is no angle offset), then the wake
cannot exert a negative torque on the bar. This angle is plotted in the left
panel of Fig.~\ref{fig:wake}, which shows that the angle offset is larger in the
\Nbody{} case than in the \SMUGGLE{} case by about a factor of three. The
center and right panels of Fig.~\ref{fig:wake} show the halo wake with respect
to the location of the bar in the \Nbody{} (center) and \SMUGGLE{} (right) cases
at one point in time. Note that in Fig.~\ref{fig:wake} we have removed the
halo material trapped in the bar, which exerts no net torque on the bar.

The presence of the gas can arrest the process by which additional material in
the dark matter halo can contribute to a wake. However, this does not explain
why the pattern speed in the \SMUGGLE{} case is nearly constant over several
Gyr. Naively, it would be a coincidence that the bar pattern speed remains
constant in the \SMUGGLE{} case, resulting from a chance cancellation of the
halo and gas torques. However, a constant pattern speed in the presence of gas
has been observed in a few simulations of barred galaxies with gas
\citep{1993AA...268...65F, 2007ApJ...666..189B, 2009ApJ...707..218V,
2010ApJ...719.1470V, 2014MNRAS.438L..81A}.

\citet{1993AA...268...65F} argue this behavior is due to the steepening of the
circular velocity curve in the central region as the bar drives gas to the
center. \citet{2009ApJ...707..218V} argue this behavior occurs when the
corotation resonance is larger than the disk radius, but we observe the behavior
when the corotation radius is well within the disk.

% Previous work has argued this is due to the bar torquing
% gas inwards, but no explanation has been given for why it might remain constant.

We propose that an equilibrium mechanism is responsible for the pattern speed
remaining approximately constant. In this scenario, residual negative torque
from the dark matter halo balances out the positive torque from the gas phase.
It has been shown when an analytic bar is forced to rotate at a constant pattern
speed for a few Gyr, the halo exerts almost no torque on the bar
\citep{2022MNRAS.513..768C}. We saw in Fig.~\ref{fig:prop} that the dark matter
halo in our simulation is still able to support some negative torque over a
several Gyr time span.

We argue that the following occurs. First, the bar is not able to slow down
quickly enough due to the positive torque of the infalling gas. This causes the
resonant halo phase space at a particular pattern speed,
$\Omega_{\textrm{p},0}$, to become mixed and no longer able to support a
negative torque.\footnote{In the halo wake picture, this corresponds to the wake
becoming fully aligned with the bar.} Second, the gas is still exerting a
positive torque on the bar, and therefore the pattern speed will again increase.
Since at higher pattern speeds the halo has not yet been totally mixed,
the halo will once again be able to exert a negative torque on the bar. The
pattern speed will then settle at a new value slightly higher than
$\Omega_{\textrm{p},0}$ where the gas and halo torques cancel. Over time, the
pattern speed should slowly increase.

\subsection{Delayed Gas Injection}
\begin{figure}
    \centering
    \includegraphics[width=9cm]{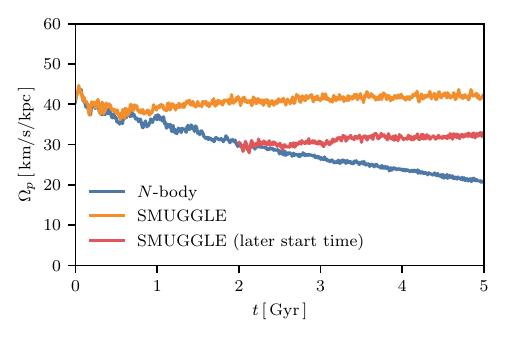}
    \caption{Pattern speed evolution with a lower initial pattern speed. We
    tested the evolution of our system when gas is added to the \Nbody{} run at
    a later time, but with all other simulation parameters kept the same. The
    setup is therefore identical to our previous runs just with a bar that is
    larger, stronger, and with a lower pattern speed. We find that the pattern
    speed evolution is very similar to our fiducial case, except that the bar
    retains its original pattern speed. This indicates a mechanism which keeps
    the bar at its formation pattern speed, and that there is not a particular
    pattern speed which the system tends to.}
    \label{fig:snap700}
\end{figure}

A clear prediction of our proposed mechanism is that the constant pattern speed
a particular galaxy will end up is somewhat arbitrary. In the real universe for
an isolated galaxy it would be the formation pattern speed of the bar while in
our simulation it is the pattern speed of the bar when gas is added to the
system. We tested this by adding gas to the system at a later time when the bar
has further grown and slowed down with time. In our particular test, we added
the gas at a time when the pattern speed is
$\sim30\,\textrm{km}/\textrm{s}/\textrm{kpc}$. As shown in
Fig.~\ref{fig:snap700}, we find that the pattern speed evolution is very similar
between the two cases (orange and red lines). If anything, the system with a
lower pattern speed seems to speed up more, which is consistent with our picture
since the stronger bar should experience a larger torque from the gas as it is
more efficient at driving gas inflows. We also show in the
Appendix~\ref{app:varyps} that more slowly rotating bars at fixed bar strength
are more efficient at driving gas inflows as well. Nonetheless, when the initial
pattern speed is lower (red line), the addition of gas does not cause the
pattern speed to quickly return to the higher value of our fiducial simulation
(orange line).

\subsection{Varying Initial Gas Fractions}
\begin{figure}
    \centering
    \includegraphics[width=9cm]{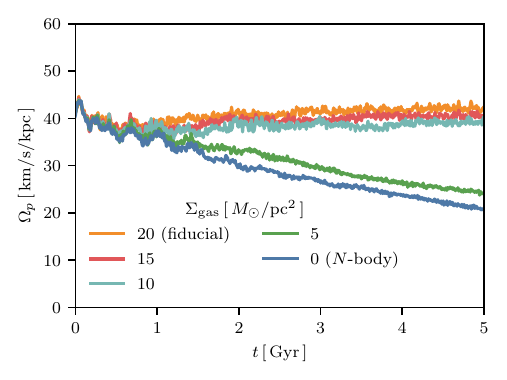}
    \caption{Pattern speed evolution with varying gas fractions. We explored the
    impact of lowering the initial gas surface density of our fiducial disk on
    the evolution of the pattern speed. The surface densities we tested of $20$,
    $15$, $10$, and $5\,\Msun/\textrm{pc}^2$ correspond to initial gas
    fractions of $16\%$, $10\%$, $7\%$, and $4\%$. We find that initial surface
    densities $20$, $15$, and $10\,\Msun/\textrm{pc}^2$ result in bars which
    remain at a constant pattern speed, while an initial surface density of
    $5\,\Msun/\textrm{pc}^2$ results in a bar which slows down in roughly the
    same manner as the \Nbody{} case.}
    \label{fig:fgas}
\end{figure}

We performed a test in which we varied the initial gas fraction of the
disk. In our fiducial run, we set the surface density of the gas disk from
$4\,\textrm{kpc}$ to $\sim9.3\,\textrm{kpc}$ to be $20\,\Msun/\textrm{pc}^2$.
We also ran with surface densities of $15$, $10$, and
$5\,\Msun/\textrm{pc}^2$. These correspond to initial gas fractions of
approximately $16\%$, $10\%$, $7\%$, and $4\%$. The pattern speed evolution is
shown in Fig.~\ref{fig:fgas}. We find that the bar in disks with
initial surface densities of $20$, $15$, and $10\,\Msun/\textrm{pc}^2$ evolve
with a constant pattern speed while the bar in a disk with initial surface
density of $5\,\Msun/\textrm{pc}^2$ slows down at a similar rate to the \Nbody{} case.

After $5\,\textrm{Gyr}$, the $10\,\Msun/\textrm{pc}^2$ simulation has a gas
fraction of $5.7\%$, but still exhibits constant pattern speed behavior. As a
result, we conclude that for the disk, bar, and halo properties considered in
this work, a gas fraction of only approximately $5\%$ is necessary in order for
the proposed stabilizing mechanism to operate. We stress that this gas
fraction threshold is only for the system considered in this work. Systems with
different structural parameters may require a different threshold. For example,
\citet{2015MNRAS.454.3166A} find slowdown behavior in a barred disk $10$ times
less massive than ours with a gas fraction of $10\%$. To make things more
complicated, \citet{2010ApJ...719.1470V} find that the gas fraction cutoff
varies with the softening length used. Determining exactly how the gas fraction
cutoff varies with these considerations deserves further attention.

\subsection{Semi-Analytic Model}
We also developed a simple semi-analytic model of a bar-disk-halo system. This
exercise demonstrates that our proposed mechanism follows from a few simple
assumptions. Our method follows closely the one developed in
\citet{2022MNRAS.513..768C}. We model the bar-disk-halo system with three
components: a dark matter \citet{1990ApJ...356..359H} halo, a
\citet{1975PASJ...27..533M} disk, and a pure quadrupole bar
described in \citet{2022MNRAS.513..768C}. The bar and disk components are just
described by their potential, but we integrate the trajectories of test
particles drawn from a Hernquist halo. Note that we do not include the
interactions between these test particles, and so this model is not
self-consistent. We give our chosen parameter values in Appendix~\ref{app:sam}.

We allow the bar in this model to rotate as a solid body. However, we crucially
allow the pattern speed of the bar to freely change with time in accordance with
the torque exerted on the dark matter halo by the bar. In particular, we
subtract the $z$-component of this torque divided by the moment of inertia of
the bar from the pattern speed at each timestep. Since the radius of corotation
\RCR{} is a parameter in the bar model from \citet{2022MNRAS.513..768C}, we
allow the moment of inertia of the bar to vary with ${\RCR}^2$. This is
inspired by the fact that the moment of inertia of an ellipsoid scales with the
sum of the square of its axes. To be more precise, we allow
\begin{equation}
I = \frac{I_6}{10^{10} \Msun (\kms)^2} \left( \frac{\RCR}{6\,\textrm{kpc}} \right)^2\textrm{,}
\end{equation}
where $I_6$ is a free parameter chosen by the user. We found that allowing
$I_6=8$ is a good approximation to our fiducial disk model. In code units, the
moment of inertia of the \SMUGGLE{} bar (i.e., the particles classified as being in
the bar) is about $2$. This is a factor of $4$ smaller than our fiducial value
of $I_6=8$, but this is probably due either to the fact that the bar does not
really rotate as a solid body or that resonantly captured stars contribute to
the real bar's effective moment of inertia \citep{1985MNRAS.213..451W}.

In addition to $I_6$, we allowed for another free parameter - the torque from
the gas phase on the bar, $\tau_{\textrm{gas}}$. This torque is applied to the
bar in the same way as the torque from the halo is applied. The torque is given
in code units ($10^{10}\Msun \left(\kms\right)^2$).

\begin{figure}
    \centering
    \includegraphics[width=\columnwidth]{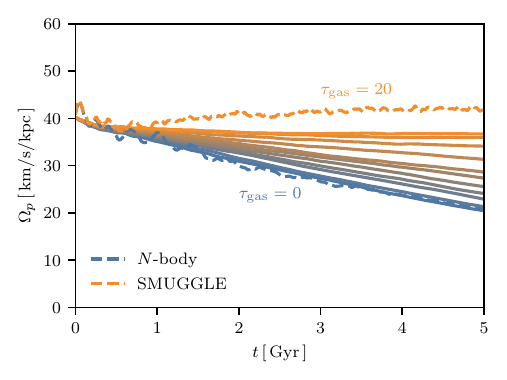}
    \caption{A comparison between the pattern speeds of our fiducial disk
    systems and a semi-analytic model. The solid lines indicate the pattern
    speeds assuming a constant positive torque, varying in increments of $2$
    from $0$ to $20$. The dashed lines indicate the pattern speed evolution from
    our fully self-consistent simulations from earlier. We find excellent
    agreement between our fiducial simulations and our semi-analytic model of a
    bar-disk-halo system. Torques are given in code units ($10^{10}\Msun
    \left(\kms\right)^2$).}
    \label{fig:sam}
\end{figure}

We show the effect of varying the gas torque $\tau_{\textrm{gas}}$ from $0$ to
$20$ in increments of $2$ in Fig.~\ref{fig:sam}. The solid lines indicate the
semi-analytic model, while the two dashed lines correspond to our fiducial
simulations introduced earlier. For reference, the average torque exerted by the
gas phase on the bar in our fiducial simulation was $11.7$ in code units. We see
in Fig.~\ref{fig:sam} that we can reproduce the stability of our fiducial gas
disk (i.e., its lack of secular evolution) simply by including a positive torque
on the order of $6$. Our semi-analytic model with no gas torque can reproduce
the pattern speed evolution of the $N$-body case.

\begin{figure}
    \centering
    \includegraphics[width=\columnwidth]{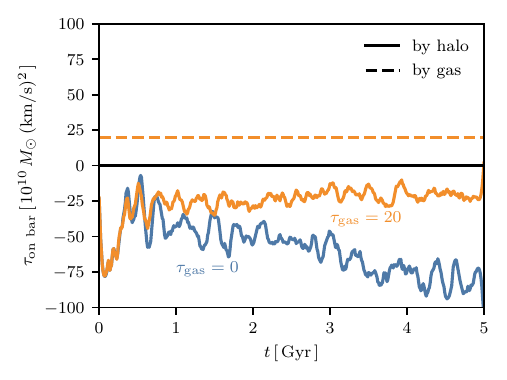}
    \caption{The torque exerted on the bar by various components in our
    semi-analytic model. Solid lines indicate the torque by the halo while the
    dashed line indicates the torque exerted by the gas phase. We chose two
    models with $\tau_{\textrm{gas}}=0$ and $20$, which most closely resemble
    our $N$-body and \SMUGGLE{} disks, respectively. This figure ought to be
    compared to the lower panel of Fig.~\ref{fig:prop}. Overall, we find good
    qualitative agreement.}
    \label{fig:sam-torque}
\end{figure}

We next take the $\tau_{\textrm{gas}}=0$ and $20$ cases from Fig.~\ref{fig:sam}
and plot the halo torque evolution. This result, given in
Fig.~\ref{fig:sam-torque}, is comparable to the lower panel of
Fig.~\ref{fig:prop}. We find that the $\tau_{\textrm{gas}}=0$ case compares
favorably to the \Nbody{} case described in previous sections. The bar exerts a
steady negative torque in this case (blue line). When a gas torque is included
(orange lines), we find that the halo's torque becomes much weaker, similar to
what we found in the \SMUGGLE{} case. The gas applies a steady positive torque
by construction. Therefore, the locking process is reproduced in our simple
semi-analytic model. Curiously, we do not see the bimodal behavior in
Fig.~\ref{fig:fgas} and described in \citet{2010ApJ...719.1470V}. It is not
presently clear why this is the case.

\subsection{Observations}
\label{ssec:observations}
Observational estimates of the pattern speeds of bars indicate that nearly all
galaxies have $1 < \Rot < 1.4$ \citep{2011MSAIS..18...23C, 2015AA...576A.102A,
2019MNRAS.482.1733G, 2020MNRAS.491.3655G}, where $\Rot$ was defined in
Section~\ref{sec:intro} to be $\Rot\equiv \RCR/\Rb$. This observational fact has
long been in conflict with the theoretical expectation that bars should slow
down, increasing \Rot{} \citep[e.g.][]{1984MNRAS.209..729T, 1985MNRAS.213..451W,
2000ApJ...543..704D}. Explanations for this discrepancy have been given in the
past. Some have argued that perhaps the central regions of dark matter haloes
are less dense than we expected from $\Lambda\textrm{CDM}$
\citep[e.g.][]{2000ApJ...543..704D,2021AA...650L..16F}. Some have argued that
perhaps bars are recurrent, short-lived phenomena, and that all the bars we see
in the local universe are very young \citep{2002AA...392...83B,
2005MNRAS.364L..18B}. Some have argued that modifications to General Relativity
ease the tension between the observed universe and $\Lambda\textrm{CDM}$
\citep[e.g.][]{2021MNRAS.503.2833R, 2021MNRAS.508..926R}.

Because such a small gas fraction is necessary for our stabilizing mechanism to
operate ($5\%$ in our Milky Way-like disk), we argue that most galaxies host a
bar that is not slowing down. This naturally explains why most observed bars are
fast rotators. However, we acknowledge two instances of reported discrepancies
between our mechanism and observations.

First, we note that \citet{2020MNRAS.491.3655G} found that the rotation
parameter \Rot{} positively correlates with gas fraction, such that galaxies
with higher gas fractions are rotating more slowly. However, it is not obvious
this is in tension with our result since the gas fraction of galaxies correlates
with other galactic properties \citep{2009ARAA..47..159B}. Furthermore, the
measurement of pattern speeds is a delicate process still prone to large errors.

Second, the work of \citet{2021MNRAS.500.4710C} and \citet{2021MNRAS.505.2412C}
have made indirect measurements of the deceleration of the bar's pattern speed
from kinematics and chemistry. We point out that these reported measurements are
not direct measurements of the Milky Way bar's deceleration. For instance,
\citet{2021MNRAS.500.4710C} measures the pattern speed based on the asymmetry of
the Hercules stream, but this can also be produced by spiral arms
\citep{2018MNRAS.481.3794H}. Much like the simulations in the present work, the
simulations of these two works do not properly account for the complicated
formation process of the Galactic bar, which may leave imprints on the present
day distribution of stars in spatial, kinematic, and chemical space. More
investigation is necessary to reconcile the present work with these two
well-executed manuscripts.

Lenticular galaxies which lack a signficant gas phase offer an opportunity
to find slowly rotating bars. It would still take several Gyr for a galaxy
hosting a fast bar to transition to the slow bar regime, so slow bars should
only occupy lenticular galaxies which have been lenticular for some time.
NGC~4277 is one such example, whose bar has been found to rotate with
$\Rot{}\sim1.8$ \citep{2022AA...664L..10B}. On the other hand, NGC~4264 has a
fast bar with $\Rot{}\sim1$ \citep{2019MNRAS.488.4972C}. The difference has been
explained by differences in the dark matter content of the galaxies
\citep{2023MNRAS.521.2227B}. We offer another explanation based on the timing of
when gas was stripped from these galaxies.

There are further examples of gas-rich galaxies hosting slow bars. For
example, UGC~628 \citep[$\Rot\sim2$;][]{2009AA...499L..25C} and NGC~2915
\citep[$\Rot>1.7$;][]{1999AJ....118.2158B}. UGC~628
has been studied in detail by \citet{2016MNRAS.463.1751C}, who note that it
indeed has a low gas fraction for galaxies of its type. NGC~2915 has a gas
fraction of $70\%$ \citep{2010ApJ...715..656W}, which would seem to be in
conflict with our prediction that only a $5\%$ gas fraction is necessary to
arrest the halo slowdown process. However, NGC~2915 has significantly different
structural properties than the Milky Way-like model we considered in this work.
In particular, it has a signficantly lower mass ($\sim10^9\Msun$ compared to
$4.8\times10^{10}\Msun$ in our model). Further work is necessary to see how the
gas fraction threshold varies with galactic properties.

\citet{2020MNRAS.495.4158F} find that quenched galaxies tend to host longer bars
than star-forming galaxies. This provides some support for our proposed
mechanism since there is evidence quenching can occur through gas depletion
\citep[e.g.][]{2021Natur.597..485W}. However, this correlation could be
explained simply by the fact that longer bars ought to be more efficient at
quenching their host galaxies \citep[e.g][]{2015AA...580A.116G}.

Finally, we mention the evolution of \Rot{} in our simulation. In the
$N$-body simulation, the bar forms with $\Rot\sim1.6$ at $t=0\,\textrm{Gyr}$,
which is already well within the slow bar regime. After $5\,\textrm{Gyr}$ of
evolution, \Rot{} has risen to $\sim1.9$. In the SMUGGLE simulation, the gas is
added to the $t=0\,\textrm{Gyr}$ snapshot, so it begins with $\Rot{}\sim1.6$. As
expected, after $5\,\textrm{Gyr}$ of evolution \Rot{} is still $\sim1.6$.
However, this relies on our measure of the bar length as being the maximum
radius of all orbits trapped in the bar, which is not an observationally
accessible measure of bar length. One would need to test different
observationally possible bar length estimators, such as ellipse fitting
\citep{1990MNRAS.245..130A, 1999AAS..140....1M, 2002MNRAS.330...35A,
2006AA...452...97M, 2009AA...495..491A, 2015AA...576A.102A}. This is beyond the
scope of our current work. Nonetheless, our prediction that \Rot{} is stable in
the presence of sufficient gas is robust. Assuming bars form with $\Rot\sim1$,
we predict this should remain the case with further evolution. Why the $N$-body
simulation forms a bar with $\Rot\sim1.6$ is a separate question deserving
further attention.

\subsection{Previous Idealized Simulation Work}
Substantial work has been devoted to the role of gas in bar dynamics. We
discuss this and highlight the novel aspects of the present investigation.

To our knowledge the first work on a barred galaxy with a gas component was
by \citet{1993AA...268...65F}. They found a stable pattern speed when a
dissipative component was added to the system, with a slight increase in the
pattern speed near the end of their simulation. However, their model was only
evolved for $\sim1\,\textrm{Gyr}$, so it is unclear if their bar exhibits a
stable pattern speed over several Gyr.

\citet{2007ApJ...666..189B} describe a model containing up to $8\%$ gas.
Their disk is similar to ours (though their dark matter halo is a factor of $10$
less massive). They find slowdown behavior up to a gas fraction of $8\%$, with
the finding that higher gas fractions lead to a reduced slowdown. However, for
the models which they show pattern speeds, they do not include star formation or
the removal of gas from the center of their disk. In preliminary work, we found
slowdown behavior in an adiabatic model with similar gas fractions but lacking
any method for removing gas from the central region (in our fiducial SMUGGLE
model this is achieved through star formation). We surmise that the removal of
gas from the central region is an important requisite for stable pattern speeds,
though a careful torque analysis is necessary to confirm this hypothesis.

\citet{2010ApJ...719.1470V} do find stable pattern speeds in models with
gas fractions as low as $8\%$ (depending on the force softening used).
Crucially, their model does contain a routine for removing gas from the central
region of the disk. These authors state the behavior is bimodal, with a clear
stable regime and a slowdown regime. These authors make no mention of the
process by which the halo braking process is arrested, as we propose in the
present paper.

\citet{2013MNRAS.429.1949A, 2014MNRAS.438L..81A} find stable evolution in \Rot{}
for a triaxial halo over $10\,\textrm{Gyr}$ with initial (final) gas fractions
of $100\%$ ($7\%$), $75\%$ ($6\%$), and $50\%$ ($5\%$). For a model with $20\%$
($3\%$) gas fraction they find an increasing \Rot{}. In their model with a
spherical halo they always find increasing \Rot{}, contrary to the present work.
We note there is a structural difference in their dark matter halo. They use a
cored isothermal sphere as opposed to our Hernquist halo. It is not clear to us
the impact this would have on the expected torque from the halo (i.e., the
velocity structure of their halo may allow for more efficient capture and thus
stronger torquing). The bar locking process (or something similar) which we
propose in this paper is not mentioned by these authors.

There may also be issues related to structural differences between their
bars and the bar considered in this work. In our case, we allowed the bar to
form in an $N$-body run and then added gas after the bar formation. In
\citet{2013MNRAS.429.1949A, 2014MNRAS.438L..81A} the bar forms from a disk that
is initially gas-rich. Neither approach is inherently better, but it is known
that in the latter case the resultant bar strength is weaker for initially
gas-rich systems \citep[e.g.,][]{2013MNRAS.429.1949A}. Weaker bars are less
efficient at driving gas inwards \citep{2004ApJ...600..595R} and thus should
experience less positive torque from the gas phase. A direct comparison based on
the torque by the gas phase on the bar is necessary.

\citet{2015MNRAS.454.3166A} describe a model containing gas and stars which
slows down with time. In their model gas is added to the system to target a gas
fraction of $10\%$. However, their disk is about a factor of $10$ less massive
than the disk considered in this work. The necessary gas fraction for a stable
pattern speed probably depends on galaxy properties like mass. It is unclear
whether $10\%$ is sufficient for their bar to have a stable pattern speed.

\subsection{Cosmological Simulations}
Barred galaxies in cosmological simulations of galaxy formation continue to be
in conflict with observations by producing bars which rotate too slowly
\citep{2017MNRAS.469.1054A, 2019MNRAS.483.2721P, 2021AA...650L..16F,
2022ApJ...940...61F}.\footnote{Though see \citet{2022ApJ...940...61F} who argue
bars have consistent pattern speeds with observations, but are too short.}
These works examine bars in EAGLE \citep{2015MNRAS.450.1937C,
2015MNRAS.446..521S}, Illustris \citep{2014Natur.509..177V,
2014MNRAS.444.1518V}, and Illustris TNG50 \citep{2019MNRAS.490.3196P,
2019MNRAS.490.3234N}. \citet{2015PASJ...67...63O} describe cosmological zoom
simulations of two barred Milky Way-like galaxies that both slow down over time.
As pointed out by \citet{2010ApJ...719.1470V}, the gas fraction cutoff for the
stable pattern speed regime increases with lower softening lengths.
\citet{2015PASJ...67...63O} use softening lengths larger than ours by about a
factor of $6$. This highlights the importance of future work exploring precisely
when barred galaxies ought to be in the stable regime.

\citet{2021AA...650L..16F} explored the evolution of bars in the Auriga
cosmological zoom simulations \citep{2017MNRAS.467..179G}. They find \Rot{}
values consistent with observations. Furthermore, their pattern speed evolution
shows some apparent periods of stability (see their Fig.~B1). One galaxy, Au26,
appears to even transition from the stable pattern speed regime to the slowing
down regime at a lookback time of $\sim1.8\,\textrm{Gyr}$.

Furthermore, the pattern speeds of bars in both cosmological simulations and the
real universe can be affected by environmental processes not included in our
simulation -- e.g., satellite infall \citep{2011Natur.477..301P}, non-sphericity
\citep{2013MNRAS.429.1949A}, rotation in the dark matter halo
\citep{2013MNRAS.434.1287S, 2014ApJ...783L..18L, 2018MNRAS.476.1331C,
2019MNRAS.488.5788C}, or perhaps even the gaseous circumgalactic medium.
Naturally, extending our present work to account for such effects is a crucial
next step in understanding the formation and evolution of galactic bars. We are
presently engaged in such an exploration.

\section{Conclusions}
\label{sec:conclusions}
We performed a simulation of a Milky Way-like galactic disk hosting a strong bar
with a state-of-the-art model for the interstellar medium. We found that the
pattern speed of the bar in this simulation does not slow down but rather
remains at a stable, constant pattern speed. We provided a simple semi-analytic
model which reproduces many of the features from our fiducial disk model.

The implications of our findings are numerous. First, we naturally explain why
nearly all observed galaxies are fast rotators without requiring the inner
regions of dark matter halos to be underdense \citep{1998ApJ...493L...5D,
2000ApJ...543..704D} or requiring new physics \citep{2021MNRAS.503.2833R,
2021MNRAS.508..926R}. Second, we show that the role of gas is of paramount
importance in studies which attempt to uncover the nature of dark matter from
its effect of slowing down the bar \citep{2021MNRAS.500.4710C,
2021MNRAS.505.2412C}. Third, we provide an explanation for how the Milky Way's
bar could be both long-lived and a fast rotator, of which there is some
observational evidence \citep{2019MNRAS.490.4740B}. And finally, we complicate
the picture of stellar radial mixing expected to sculpt the Milky Way's
disk \citep{2012MNRAS.420..913B, 2015ApJ...808..132H}, a process which relies
upon the pattern speed of the bar to change with time. The radial mixing of
the gas phase induced by the bar, as predicted in \citet{2011MNRAS.415.1027H},
might have implications for the radial metallicity gradients of galaxies. Our
work does not alter expectations for radial mixing induced by spiral arms
\citep{2002MNRAS.336..785S}.

We found that below a certain gas fraction, bars should still be able to slow
down. Therefore, we expect barred spiral galaxies which have been gas-poor for
extended periods of time to be rotating very slowly. We therefore predict that
observations which target such galaxies (e.g., lenticular barred galaxies
\citep{2009ARAA..47..159B}) would find slowly rotating bars.\footnote{This does
not mean that we predict \textit{all} gas-poor galaxies should be slowly
rotating. Indeed, they would need to be gas-deficient for several $\textrm{Gyr}$
before they would be classified as slow rotators.} There does exist
examples of galaxies known to be slow rotators -- the low surface brightness
galaxy UGC~628 \citep{2009AA...499L..25C}, lenticular galaxy NGC~4277
\citep{2022AA...664L..10B}, and NGC~2915 \citep{1999AJ....118.2158B}. UGC~628
has been studied in detail by \citet{2016MNRAS.463.1751C}, who note that it
indeed has a low gas fraction for galaxies of its type. NGC~2915 has a gas
fraction of $70\%$ \citep{2010ApJ...715..656W}, which would seem to be in
conflict with our prediction that only a $5\%$ gas fraction is necessary to
arrest the halo slowdown process. However, NGC~2915 has significantly different
structural properties than the Milky Way-like model we considered in this work,
discussed in Section~\ref{ssec:observations}. We predict a general trend that
bars in gas-rich spiral galaxies should rotate quickly while some bars in
gas-poor spiral galaxies should rotate slowly.

Snapshots at $500\,\textrm{Myr}$ cadence are publicly available at
\url{https://drive.google.com/drive/folders/1nZF7mZ98T0QPc2pt7CVCUUdQAm7Q9nK-?usp=sharing}.

\begin{acknowledgments}
  We would like to thank Greg~L. Bryan, Neal~J. Evans, Drummond~B. Fielding, Keith
  Hawkins, Jason~A.~S. Hunt, Sarah~M.~R. Jeffreson, Kathryn~V. Johnston,
  Peter~M.~W. Kalberla, Jürgen Kerp, Julio~F. Navarro, Dylan Nelson, Suchira
  Sarkar, Joshua~S. Speagle, Martin~D. Weinberg, and Yanfei Zou for helpful
  discussions. A.B. would like to thank Todd Phillips for helpful discussions.
  Some computations in this paper were run on the FASRC Cannon cluster supported
  by the FAS Division of Science Research Computing Group at Harvard University.
  Resources supporting this work were also provided by the NASA High-End
  Computing (HEC) Program through the NASA Advanced Supercomputing (NAS)
  Division at Ames Research Center. This research has made use of the Spanish
  Virtual Observatory (https://svo.cab.inta-csic.es) project funded by
  MCIN/AEI/10.13039/501100011033/ through grant PID2020-112949GB-I00. A.B. was
  supported by the Future Investigators in NASA Earth and Space Science and
  Technology (FINESST) award number 80NSSC20K1536 during the completion of this
  work. E.D. was partially supported by HST grants HST-AR-16363.001 and
  HST-AR-16602.006-A and by NASA Award NASA 80NSSC22K0761. J.Q. acknowledges
  support from NSF grant AST-2008490. L.V.S. is grateful for financial support
  from NASA ATP 80NSSC20K0566, NSF AST 1817233 and NSF CAREER 1945310 grants.
  P.T. acknowledges support from NSF grant AST-1909933, AST-2008490, and NASA
  ATP Grant 80NSSC20K0502. M.V. acknowledges support through NASA ATP
  19-ATP19-0019, 19-ATP19-0020, 19-ATP19-0167, and NSF grants AST-1814053,
  AST-1814259, AST-1909831, AST-2007355 and AST-2107724.
\end{acknowledgments}

\software{
{\sc agama} \url{https://github.com/GalacticDynamics-Oxford/Agama}, {\sc
astropy} \citep{astropy:2013, astropy:2018}, {\sc h5py}
\url{http://www.h5py.org/}, {\sc inspector\_gadget}
\url{https://bitbucket.org/abauer/inspector_gadget/}, {\sc joblib}
\url{https://joblib.readthedocs.io/en/latest/}, {\sc matplotlib}
\citep{Hunter:2007}, {\sc numba} \citep{lam2015numba}, {\sc numpy}
\citep{harris2020array}, {\sc scipy} \citep{2020SciPy-NMeth}, {\sc tqdm}
\url{https://tqdm.github.io/}
          }

\appendix

\section{Bar Decomposition}
\label{app:bardecomp}
\begin{figure*}
    \centering
    \includegraphics[width=18cm]{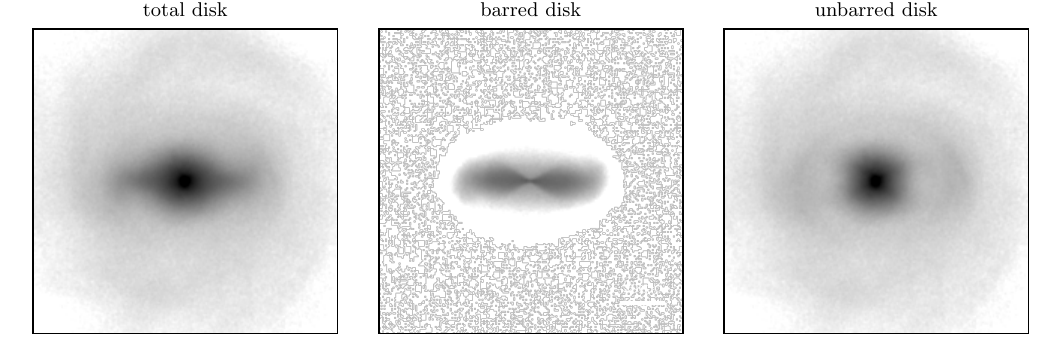}
    \caption{Disk decomposition into the barred and unbarred disk. This
    procedure is based on \citet{2016MNRAS.463.1952P}. The \textit{left panel}
    shows a face-on surface density projection through the stellar component of
    the \SMUGGLE{} simulation (disk and bulge) at $t=1\,\textrm{Gyr}$. The
    \textit{middle panel} shows the component of the disk identified as being
    trapped in the bar while the \textit{right panel} shows the component of the
    disk identified as not being trapped in the bar. The fact that the untrapped
    stars form a roughly axisymmetric structure indicates our bar decomposition
    is sufficiently accurate. We have computed that $76\%$ of the second Fourier
    component resides in the stars classified as being trapped in the bar.}
    \label{fig:decomp}
\end{figure*}

Computing the length of the bar and the torque on the bar by different
components requires us to decompose the disk into a component which is trapped
by the bar and a component which is untrapped. In order to do this, we follow
closely the technique developed in \citet{2016MNRAS.463.1952P}. We analyzed the
orbit of each star particle (meaning initial disk, bulge, and newly formed
stars) by extracting the $x$-$y$ positions of the apoapse of each in a frame
corotating with the bar, where apoapses are defined as local maxima in $r$. For
each apoapse, we searched for the $19$ closest apoapses in time and applied a
$k$-means clustering algorithm on this set of $20$ points with $k=2$. We then
computed for each of the two clusters the average angle from the bar
$\left<\Delta \phi\right>_{0,1}$, the standard deviation in $R$ of the points
${\sigma_R}_{0,1}$, and the average radius of the cluster
$\left<R\right>_{0,1}$. At each apoapse, a particle was considered to be in the
bar if it met the following criteria:
\begin{equation}
\textrm{max}\left(\left<\Delta \phi\right>_{0,1}\right) < \pi / 8
\end{equation}
\begin{equation}
\frac{{\sigma_R}_0 + {\sigma_R}_1}{\left<R\right>_0 + \left<R\right>_1} < 0.22
\end{equation}
These criterion are slightly different and simplified from the ones used in
\citet{2016MNRAS.463.1952P}, but we found them to empirically work well at
decomposing the disk into a bar and disk component. In Fig.~\ref{fig:decomp}, we
show an example of this decomposition. The \textit{left} panel shows a surface
density projection of the stellar disk and bulge (including newly formed stars)
from the \SMUGGLE{} model after $1\,\text{Gyr}$ of evolution in a frame such that
the bar is aligned with the $x$-axis. The \textit{middle} panel shows a
projection of the subset of stars that are identified as being trapped in the
bar and the \textit{right} panel shows a projection of the stars that are not
identified as being trapped. The fact that the \textit{right} panel is roughly
axisymmetric indicates the bar decomposition is performing adequately.

We computed the second Fourier component $A_2$ for all particles classified as
barred and unbarred. We found that $76\%$ of the total $m=2$ Fourier component
is in the particles classified as barred (i.e.,
$A_{2,\textrm{bar}}/A_{2,\textrm{tot}}\sim0.76$). Some of this is probably
coming from the $m=4$ component (e.g., boxy orbits) being classified as
unbarred, or the presence of weak spiral arms. See also
\citet{2021MNRAS.500..838P} for more details on the orbit family breakdown.

\section{Varying Pattern Speed}
\label{app:varyps}
When the bar slows down, we argue that this induces a larger positive torque
from the gas phase. Only gas within corotation will flow inwards, while gas
outside corotation will flow outwards \citep{2011MNRAS.415.1027H}. Since the
corotation radius is larger for more slowly rotating bars, it follows that
more slowly rotating bars should be more efficient at driving gas inflows and
thus experience a larger positive torque from the gas phase.

\begin{figure*}
    \centering
    \includegraphics[width=9cm]{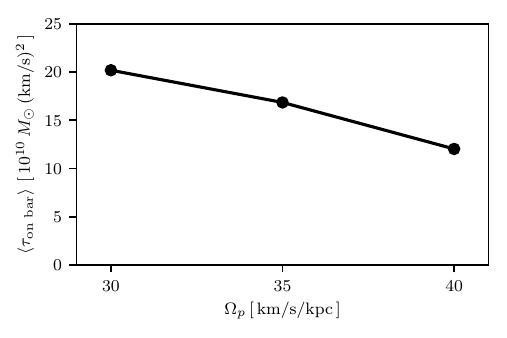}
    \caption{Average torque exerted by gas on a bar which rotates at a fixed
    pattern speed. Since only gas within the corotation radius is able to infall
    and slower bars have larger corotation radii, slower bars experience a
    larger net torque than faster bars. The setup of the simulations used here
    is identical to the \SMUGGLE{} case discussed earlier, except the \Nbody{} disk
    is rotated as a solid body with a constant angular
    velocity.}
    \label{fig:equil}
\end{figure*}

We performed an experiment to test this hypothesis by freezing the stellar
disk in the \SMUGGLE{} run and forcing it to rotate at a constant angular rate.
This has the effect of forcing the bar to rotate as a solid body at a constant
angular rate which we control. The gas is evolved self-consistently with this
rotating disk. We measured the torque on the bar by the gas phase at different
rotation rates. The result of this experiment is illustrated in
Fig.~\ref{fig:equil}, which shows that a more slowly rotating bar experiences a
larger positive torque from the gas.

We also note that since \citet{2011MNRAS.415.1027H} predicts gas outside of
corotation will flow outward, the bar should exert a positive torque on that
gas. Indeed, we measured the average torque on gas outside corotation from
$t=3\,\textrm{Gyr}$ to $5\,\textrm{Gyr}$ to be $0.87$ in code units
($10^{10}\,M_{\odot}\,(\text{km}/\text{s})^2$). For reference, the average
torque inside corotation is $-10.8$ over the same time period and in the same
units. So, while gas outside corotation does experience a positive torque, the
total torque on the gas phase is still negative.

\begin{figure*}
    \centering
    \includegraphics[width=9cm]{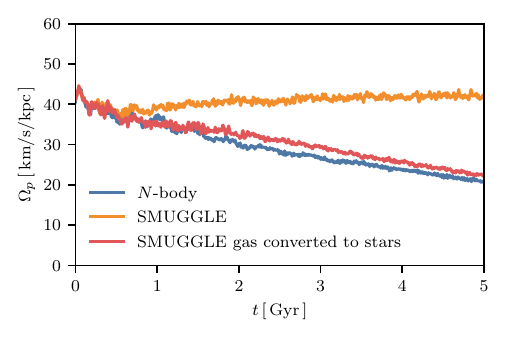}
    \caption{Pattern speed evolution of a model in which we instaneously
    add stars instead of gas to the simulation, with the same density profile as
    the gas phase. The pattern speed evolution in this case is qualitatively
    similar to that of the $N$-body case, with a slight offset in the pattern
    speed. This test demonstrates that the stable pattern speed evolution in the
    SMUGGLE case is not simply a consequence of the change in potential imposed
    in our initial conditions.}
\label{fig:ps-star}
\end{figure*}

\section{Stars Instead of Gas}
In the SMUGGLE model considered in this work, we instantaneously added gas to
the $N$-body system after $1.5\,\textrm{Gyr}$ of evolution. One might wonder if
this sudden change to the potential is responsible for the stable pattern speed
evolution. To test whether this is the case, we added mass to the system in the
same way we did for the SMUGGLE model, but using collisionless particles instead
of gas. The result of this experiment is shown in Fig.~\ref{fig:ps-star}. While
there is an offset compared to the pure $N$-body case, we see that the pattern
speed evolution is broadly consistent with a declining pattern speed. This
indicates that the gas phase is responsible for the stable pattern speed.

\begin{figure}
    \centering
    \includegraphics[width=9cm]{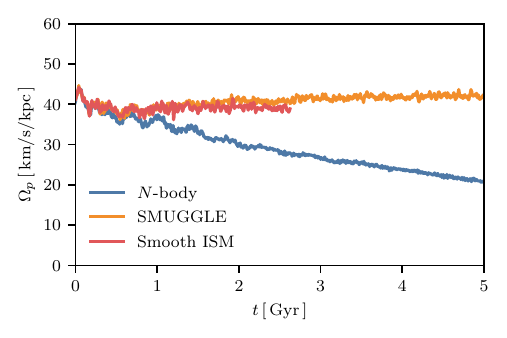}
    \caption{Pattern speed evolution of a smooth ISM model. This evolution is
    shown for the fiducial disk in the \Nbody{} (blue), \SMUGGLE{} (orange), and
    smooth ISM (red) cases. The smooth ISM model is an older model for the ISM
    which treats its multiphase nature in a subgrid fashion
    \citep{2003MNRAS.339..289S}. This fundamentally differs from the \SMUGGLE{}
    model, which explicitly resolves the hot and cold phases of the ISM
    \citep{2019MNRAS.489.4233M}. The pattern speed in the smooth ISM case is
    broadly similar to the evolution in the \SMUGGLE{} case. This shows that the
    stability of the pattern speed is not simply a result of our assumed model
    for the ISM.}
\label{fig:GFM}
\end{figure}

\section{Smooth Interstellar Medium}
We performed a simulation of the same disk but with a simpler model of the
interstellar medium \citep{2003MNRAS.339..289S}, closer to standard methods used
in cosmological simulations of galaxy formation and described in more detail in
Section~\ref{sec:methods}. The result of this test is presented in
Fig.~\ref{fig:GFM}. We find that the pattern speed evolution is nearly the same
in this case, and so conclude that our result is not sensitive to the details of
the model for the interstellar medium.

\section{Semi-Analytic Model Parameters}
\label{app:sam}
Our semi-analytic model consisted of a three-component bar-disk-halo system. We
describe here the parameters we chose for these components. The parameters of
the disk and halo were chosen to match closely what we used in our fiducial
simulations. The system can thus be understood as being roughly similar to the
Milky Way, though no careful analysis has been performed to ensure the closest
match possible.

For the dark matter halo, we used a Hernquist potential
\citep{1990ApJ...356..359H} with mass $10^{12}\,\Msun$ and a scale length of
$26.2\,\textrm{kpc}$. For the stellar disk, we used a Miyamoto-Nagai disk
\citep{1975PASJ...27..533M} with mass $4.8\times10^{10}\,\Msun$, radial scale
length of $2.67\,\textrm{kpc}$, and vertical scale length of
$0.32\,\textrm{kpc}$. For the bar, we used the quadrupole potential described in
\citet{2022MNRAS.513..768C}. We used their fiducial parameter values --
specifically, we set $A=0.02$, $b=0.28$, and $v_c = 235\,\kms$. Our initial
pattern speed is always set to $40\,\kms/\textrm{kpc}$.

We integrated our model for $5\,\textrm{Gyr}$ with a timestep of
$0.01\,\textrm{Gyr}$.

\section{Comparison to the Milky Way}
\label{app:milkyway}
For several Gyr, our fiducial disk exhibits several properties in reasonable
agreement to the Milky Way. This is uncommon in models of galaxies that include
the gas phase of the disk but no circumgalactic medium. As mentioned earlier,
the pattern speed seems to match the observed pattern speed of the Milky Way's
bar \citep{2019MNRAS.490.4740B}. We briefly summarize some of the other ways our
disk is comparable to the Milky Way.

We computed the circular velocity curve of our model using the \texttt{AGAMA}
package \citep{2019MNRAS.482.1525V}. We fit the baryonic component (stellar
disk, bulge, gas, and newly formed stars) with an axisymmetric cylindrical
spline with $20$ grid points in both the radial and vertical direction spanning
$0.2$ to $50\,\textrm{kpc}$ in the radial direction and from $0.02$ to
$10\,\textrm{kpc}$ in the vertical direction. We fit the dark matter halo using
an axisymmetric multipole fit with a maximum angular harmonic coefficient
of $l=2$, to account for the compression of the halo by the disk. We plot
the circular velocity curve at $t=1\,\textrm{Gyr}$ in Fig.~\ref{fig:vcirc}
compared to observational estimates \citep{2019ApJ...871..120E}. The \SMUGGLE{}
disk (which includes additional mass in the form of gas) has a slightly higher
circular velocity than the \Nbody{} disk which, itself, is slightly higher than
the observational estimates. Overall, though, the circular velocity curves
between our model and that observed in the Milky Way are broadly consistent.

\begin{figure*}
    \centering
    \includegraphics[width=9cm]{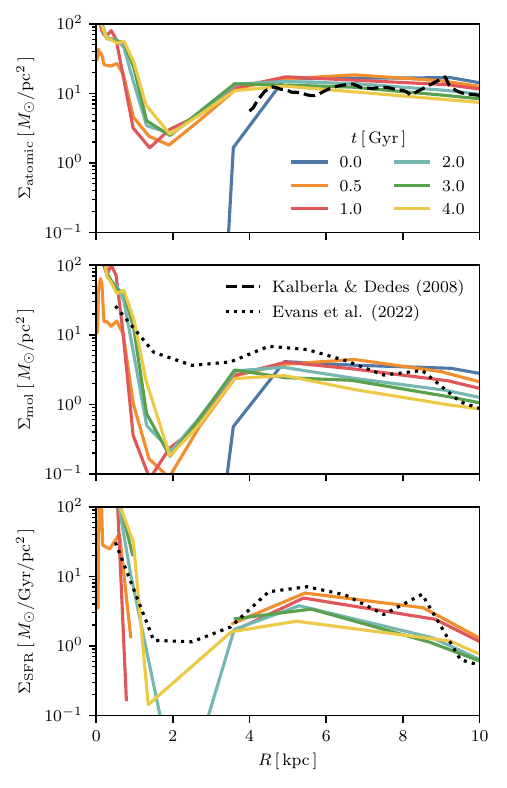}
    \caption{The time evolution of the atomic gas surface density
    (\textit{upper}), molecular gas surface density (\textit{middle}) and the
    star formation rate (SFR) surface density (\textit{lower}) at various times
    during our fiducial simulation. Colored lines indicate the profiles at
    selected times during the simulation while the black dashed lines indicate
    observations for the atomic gas \citep{2008AA...487..951K} and black dotted
    lines indicate a model which allows the CO-to-H$_2$ conversion factor
    $X_{\textrm{CO}}$ to vary with metallicity \citep{2022ApJ...929L..18E}.
    Molecular gas surface densities were provided separately (N. Evans, private
    communication). We see that the molecular gas and SFR surface densities are
    within an order of magnitude of the Milky Way's typical values at all times.
    We see a sharp decrease in the gas and SFR surface densities along the
    extent of the bar from $\sim1$ to $\sim4\,\textrm{kpc}$, related to the gas
    inflow in this region.}
    \label{fig:surf}
\end{figure*}

\begin{figure}
    \centering
    \includegraphics[width=9cm]{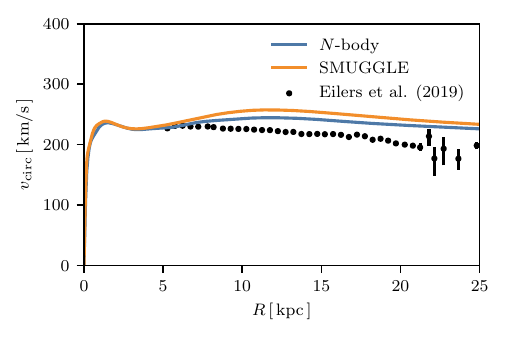}
    \caption{The circular velocity curve of our setups at $t=1\,\textrm{Gyr}$.
    This curve is shown for the \Nbody{} run (blue) and the \SMUGGLE{} run
    (orange) compared to observational estimates for the Milky Way
    \citep{2019ApJ...871..120E}. We see that the circular velocity curve for
    both runs is marginally larger than the Milky Way's, but still comparable.
    The \SMUGGLE{} circular velocity curve is larger than the \Nbody{} curve due
    to the additional mass in the gas phase.}
    \label{fig:vcirc}
\end{figure}

We also show the evolution of the surface density profile in Fig.~\ref{fig:surf}
We find that in our simulation the atomic and molecular gas surface density and
the SFR surface density is broadly consistent with the expected values for the
Milky Way \citep{2008AA...487..951K,2022ApJ...929L..18E}. The discrepancy between
$1$ and $4\,\textrm{kpc}$ in the molecular and SFR surface density is probably due
to the fact that the distances to molecular clouds which underlines this work
used a simple kinematic distance based on an axisymmetric model of the Milky
Way \citep{2017ApJ...834...57M}, which is not accurate in the bar region where gas
has large non-circular velocities.

We measured the initial scale height of the atomic gas disk in a bin
extending from $R=7.5\,\textrm{kpc}$ to $R=8.5\,\textrm{kpc}$. The initial
vertical profile is well-fit by a Gaussian with a scale height of
$110\,\textrm{pc}$. At $t=1\,\textrm{Gyr}$, the vertical profile in the same
radial bin is better fit by an exponential profile with scale height of
$74\,\textrm{pc}$. These are somewhat lower than the observed value in the HI
disk of $\sim200\,\textrm{pc}$ \citep{1995ApJ...448..138M, 2017AA...607A.106M}.
This may be caused by the model in our simulations not driving enough turbulent
pressure, and is an interesting avenue of further investigation.

%% For this sample we use BibTeX plus aasjournals.bst to generate the
%% the bibliography. The sample631.bib file was populated from ADS. To
%% get the citations to show in the compiled file do the following:
%%
%% pdflatex sample631.tex
%% bibtext sample631
%% pdflatex sample631.tex
%% pdflatex sample631.tex

\bibliography{ref}{}
\bibliographystyle{aasjournal}

%% This command is needed to show the entire author+affiliation list when
%% the collaboration and author truncation commands are used.  It has to
%% go at the end of the manuscript.
%\allauthors

%% Include this line if you are using the \added, \replaced, \deleted
%% commands to see a summary list of all changes at the end of the article.
%\listofchanges

\end{document}